\pdfoutput=1

\documentclass[11pt]{article}

\usepackage[final]{acl}

\usepackage{times}
\usepackage{latexsym}

\usepackage[T1]{fontenc}

\usepackage[utf8]{inputenc}

\usepackage{microtype}

\usepackage{inconsolata}

\usepackage{graphicx}

\usepackage{amsmath}
\usepackage{booktabs}
\usepackage{multirow}
\usepackage{makecell}
\usepackage{tcolorbox}
\newcommand{\datasetname}{\textsc{BoolQuestions}}
\newcommand{\taskname}{BDR}

\newcommand{\operator}[1]{{\color{blue} \textbf{#1}}}
\newcommand{\keyinfo}[1]{{\color{green!55!black} \textbf{\textit{#1}}}}
\newcommand{\negkeyinfo}[1]{{\color{red!80!black} \textbf{\textit{#1}}}}

\title{\datasetname: Does Dense Retrieval Understand \\ Boolean Logic in Language?}

\author{
 \textbf{Zongmeng Zhang\textsuperscript{1}},
 \textbf{Jinhua Zhu\textsuperscript{1}},
 \textbf{Wengang Zhou\textsuperscript{1,3}},
\\
 \textbf{Xiang Qi\textsuperscript{2}},
 \textbf{Peng Zhang\textsuperscript{2}},
 \textbf{Houqiang Li\textsuperscript{1,3}}
\\
 \textsuperscript{1}University of Science and Technology of China
 \textsuperscript{2}Ant Group
\\
 \textsuperscript{3}Institute of Artificial Intelligence, Hefei Comprehensive Nation Science Center
\\
\texttt{
    \{zhangzm, teslazhu\}@mail.ustc.edu.cn, }\\
\texttt{
    \{zhwg, lihq\}@ustc.edu.cn, \{qixiang.qx, minghua.zp\}@antgroup.com
}
}

\begin{document}
\maketitle
\begin{abstract}

Dense retrieval, which aims to encode the semantic information of arbitrary text into dense vector representations or embeddings, has emerged as an effective and efficient paradigm for text retrieval, consequently becoming an essential component in various natural language processing systems. These systems typically focus on optimizing the embedding space by attending to the \textit{relevance} of text pairs, while overlooking the \textit{Boolean logic} inherent in language, which may not be captured by current training objectives. In this work, we first investigate whether current retrieval systems can comprehend the Boolean logic implied in language. To answer this question, we formulate the task of Boolean Dense Retrieval and collect a benchmark dataset, \datasetname, which covers complex queries containing basic Boolean logic and corresponding annotated passages. Through extensive experimental results on the proposed task and benchmark dataset, we draw the conclusion that current dense retrieval systems do not fully understand Boolean logic in language, and there is a long way to go to improve our dense retrieval systems. Furthermore, to promote further research on enhancing the understanding of Boolean logic for language models, we explore Boolean operation on decomposed query and propose a contrastive continual training method that serves as a strong baseline for the research community.\footnote{Code and dataset are available at \url{https://github.com/zmzhang2000/boolean-dense-retrieval}.}
\end{abstract}

\section{Introduction}


\begin{figure}[!t]
    \framebox{
        \parbox{0.45\textwidth}{
            \small
            \textbf{AND Question:} How can I start a \keyinfo{career in the accounting field} \operator{and} pursue an \keyinfo{online degree program}?
            \smallskip\newline
            \textbf{OR Question:} What are the impacts of \keyinfo{global warming} \operator{or} \keyinfo{climate change} on nature and humans?
            \smallskip\newline
            \textbf{NOT Question:} What causes \keyinfo{upper abdomen pain} \operator{but is unrelated to} \negkeyinfo{liver issues}?
            \smallskip\newline
            \textbf{Question in MS MARCO:} What flower is symbol of endurance?
            \smallskip\newline
            \textbf{Question in Natural Questions:} Who sings Does He Love me with Reba?
        }
    }
    \caption{Examples of the AND, OR and NOT question in \datasetname and questions from MS MARCO and Natural Questions. Questions in \datasetname{} are more complex to understand than those in MS MARCO and Natural Questions.} \label{fig:example}
\end{figure}

Text retrieval is a fundamental component of various natural language processing systems, including question answering~\cite{chen-etal-2017-reading,karpukhin-etal-2020-dense}, dialogue systems~\cite{Chen2017SurveyDialogue}, web search~\cite{mitra-2017-learning} and so on. In the era of large language models, text retrieval has become increasingly critical as it provides an offline and incrementally updatable knowledge database, directly influencing the reliability and quality of generated responses~\cite{karpukhin-etal-2020-dense,shuster-etal-2021-retrieval-augmentation} in the Retrieval-Augmented Generation paradigm~\cite{Lewis2020rag}.

Traditional text retrieval methods estimate the relevance of query and document based on lexical overlap~\cite{salton1975vector,SALTON1988termweighting}. Utilizing the "bag-of-words" assumption and set theory~\cite{WALLER1979mathematical,bookstein1980fuzzy}, these methods organize text content in the form of inverted indexes~\cite{zobel1998inverted,zobel2006inverted}, which handle Boolean logic efficiently. The Boolean retrieval model was later extended to return ranked lists of documents by leveraging term weights~\cite{Salton1983extended}. To more flexibly consider the weights of different words, probabilistic models such as BM25~\cite{Robertson2009Probabilistic} and statistical language modeling~\cite{zhai-2007-statistical} were introduced. Nevertheless, these probabilistic models still rely on lexical overlap while omitting Boolean operations, making it hard to handle Boolean logic in queries.

With the advent of deep learning, these hand-crafted sparse text features have gradually been replaced by low-dimensional dense vectors learned by neural networks~\cite{reimers-gurevych-2019-sentence,karpukhin-etal-2020-dense,qu-etal-2021-rocketqa}, particularly using LSTM~\cite{hochreiter1997long} and the powerful Transformer architecture~\cite{Vaswani2017Attention}. Unlike sparse vector spaces, dense vectors are believed to capture the semantics implied in texts~\cite{zhao2024densesurvey}. In these frameworks, retrieval systems are expected to assign high scores to ground-truth query-document pairs and relatively lower scores to irrelevant or randomly combined pairs, often framed within modern deep learning paradigms such as contrastive learning frameworks. Benefiting from pre-trained language models, text retrieval has achieved significant performance improvements~\cite{guo2022semantic,fan2022pre,yates-etal-2021-pretrained}.

However, dense retrieval systems primarily focus on encoding the \emph{relevance} of texts, which may be insufficient for handling complex Boolean logic in natural language queries. Specifically, since current retrieval systems do not incorporate Boolean logic in their training paradigms, there is no guarantee of comparability in the output scores for query-document pairs involving Boolean logic. For example, the relevance scores for queries containing logical NOT, which exclude undesired information, have not been thoroughly investigated. In contrast, lexical-based retrieval systems effectively address complex Boolean logic using set theory, a method not directly applicable to dense retrieval. Consequently, it remains unclear whether dense retrieval systems can fully comprehend and process complex Boolean logic.

To answer this question,  we formulate the task of Boolean Dense Retrieval~(\taskname) and collect a benchmark dataset, \datasetname, which includes complex queries containing basic Boolean logic and corresponding annotated passages. We evaluate several state-of-the-art dense retrieval systems on the proposed task and benchmark dataset, finding significant performance drops on NOT questions, as illustrated in Figure~\ref{fig:example}. Our findings indicate that current dense retrieval models do not fully understand Boolean logic in language.

Based on these observations, we generate additional training data for NOT questions and propose a contrastive continual learning baseline to enhance the understanding of logical NOT in natural language. While the proposed baseline reduces the negative rate of the passage list returned by retrieval systems, it also slightly sacrifices accuracy. This phenomenon underscores the difficulty of understanding logical NOT in natural language for dense retrieval systems. Nevertheless, the proposed baseline serves as a starting point for further research on dense retrieval in the community.

\section{Related Work}
In this section, we briefly introduce several highly related works about our work.
\subsection{Sparse Retrieval}

Sparse retrieval refers to the lexical-based retrieval model that conceives both the query and documents as a set of terms, known as the bag-of-words assumption. Boolean model is the most classical model developed in the early stage of information retrieval~\cite{Stefan2016}. Queries in Boolean retrieval model are formed as Boolean expressions, which comprise terms joint by Boolean operators including ``AND'', ``OR'' and ``NOT''. The retrieval process is typically based on set theory and Boolean algebra. Inverted index~\cite{zobel1998inverted,zobel2006inverted} is utilized as the data structure to implement the Boolean model.

Due to the special form of queries, logic in queries is precisely delivered to the retrieval system. Despite that, the binary decision of Boolean model lacks the ability of providing the ranking of the documents. TF-IDF and BM25~\cite{Robertson2009Probabilistic} term weighting is then proposed to assign continuous relevance scores to documents, turning retrieval to a ranking task. However, the schema of retrieving documents barely on the ranking scores of documents discards the ability to express the Boolean logic explicitly, and thus struggles to tackle the complex Boolean logic in queries.

\subsection{Dense Retrieval}

With the re-surge of deep learning, the hand-crafted scoring function has been gradually replaced by learnable neural networks. Specifically, texts are encoded to low-dimensional dense vectors and the relevance of texts is measured in latent semantic space. LSTM~\cite{hochreiter1997long} and the powerful Transformer architecture~\cite{Vaswani2017Attention} are typically used as encoders and trained with contrastive learning frameworks that pull together the representations of relevant texts and push apart those of irrelevant texts~\cite{reimers-gurevych-2019-sentence,karpukhin-etal-2020-dense,qu-etal-2021-rocketqa}. The pre-training then fine-tuning paradigm~\cite{devlin-etal-2019-bert,liu2023pretrain} has significantly pushed forward the development of dense retrieval~\cite{guo2022semantic,fan2022pre,yates-etal-2021-pretrained}.

Despite these dedicated efforts, there is no evidence suggesting that dense retrievers truly understand logic in language. In this work, we propose the task of Boolean dense retrieval and collect a benchmark dataset to investigate whether dense retrieval models understand the Boolean logic in natural language.

\subsection{Datasets for Modern Retrieval}

A number of datasets are publicly released to provide large-scale relevance judgments or test beds for retrieval, significantly facilitating the research of modern text retrieval systems. Natural Questions~\cite{kwiatkowski-etal-2019-natural} includes questions from Google search engines, along with related paragraphs and answer spans from top-ranked Wikipedia documents. MS MARCO~\cite{Bajaj2018MS} consists of a large number of queries from Bing search logs and annotated relevant passages collected from Web pages. Variants of MS MARCO like mMARCO~\cite{bonifacio2021mmarco} and MS MARCO Chameleons~\cite{Arabzadeh2021MSMARCOChemeleons} are created to enrich the evaluation characteristics with respect to multilingual and question difficulty. Domain-specific retrieval datasets such as those for financial~\cite{Maia2018financial}, scientific~\cite{wadden-etal-2020-fact} and biomedical~\cite{Tsatsaronis2015AnOO} fields are also released to advance the development of retrieval in other areas. Additionally, BEIR~\cite{thakur2021beir} and KILT~\cite{petroni-etal-2021-kilt} aggregate representative datasets to measure the overall performance of retrieval models.

To the best of our knowledge, none of these efforts probes into the logic implied in queries. BoolQ~\cite{clark-etal-2019-boolq} contains questions that can be answered with barely ``yes'' or ``no'', but does not involve complex Boolean logic in its queries. Although \citet{malaviya-etal-2023-quest} and \citet{zhong-etal-2023-romqa} construct questions with Boolean logic, their atomic questions are entity queries rather than natural language questions. This work is the first to construct a benchmark dataset for Boolean dense retrieval.

\section{Task Definition}

In this section, we initially review the technique of dense retrieval, subsequently give a formal definition of what we term as Boolean dense retrieval.

The objective of text retrieval is to identify documents within a substantial text corpus that are relevant to specific queries. We denote the query text by $q$ and the corpus by $\mathcal{D} = \big\{{d_i}\big\}_{i=1}^{|\mathcal{D}|}$. Typically, retrieval models evaluate each document in the corpus, assigning relevance scores and returning the top $k$ documents based on these scores. The scoring function is typically implemented through either lexical or semantic matching methodologies.

In this work, we focus on dense retrieval models, which are reputed for their capacity to encode the semantics of language into a dense vector representation. In dense retrieval, both query and document texts are represented as dense vectors, and their relevance is assessed within this vector space, expressed as:
\begin{equation}
    \text{Rel}(q, d) = f_{\text{sim}}(\phi(q), \psi(d)),
\end{equation}
where $\phi(\cdot)$ and $\psi(\cdot)$ are encoding functions that transform text into dense vectors, and $f_{\text{sim}}$ is a similarity function designed to measure the distance of these vectors as an indication of relevance between the encoded texts.

Building upon the conventional dense retrieval model, we introduce the concept of Boolean dense retrieval. This approach enables the integration of complex Boolean logic within text queries. Formally, a complex Boolean query is semantically equivalent to a Boolean logic expression consisting of several simpler queries, represented as:
\begin{equation}
    q \iff q_1 \star q_2 \star \cdots \star q_m,
\end{equation}
where $q_i$ represents individual simple queries and $\star$ denote the Boolean operations chosen from $\{\text{AND}, \text{OR}, \text{NOT}\}$.

Apart from the construction of the query, the retrieval process remains consistent as outlined above. Moreover, because our query incorporates a more intricate construction method, we emphasize the need for more sophisticated evaluation metrics and training techniques for retrieval models.

\begin{figure*}[t]
  \centering
  \includegraphics[width=\linewidth]{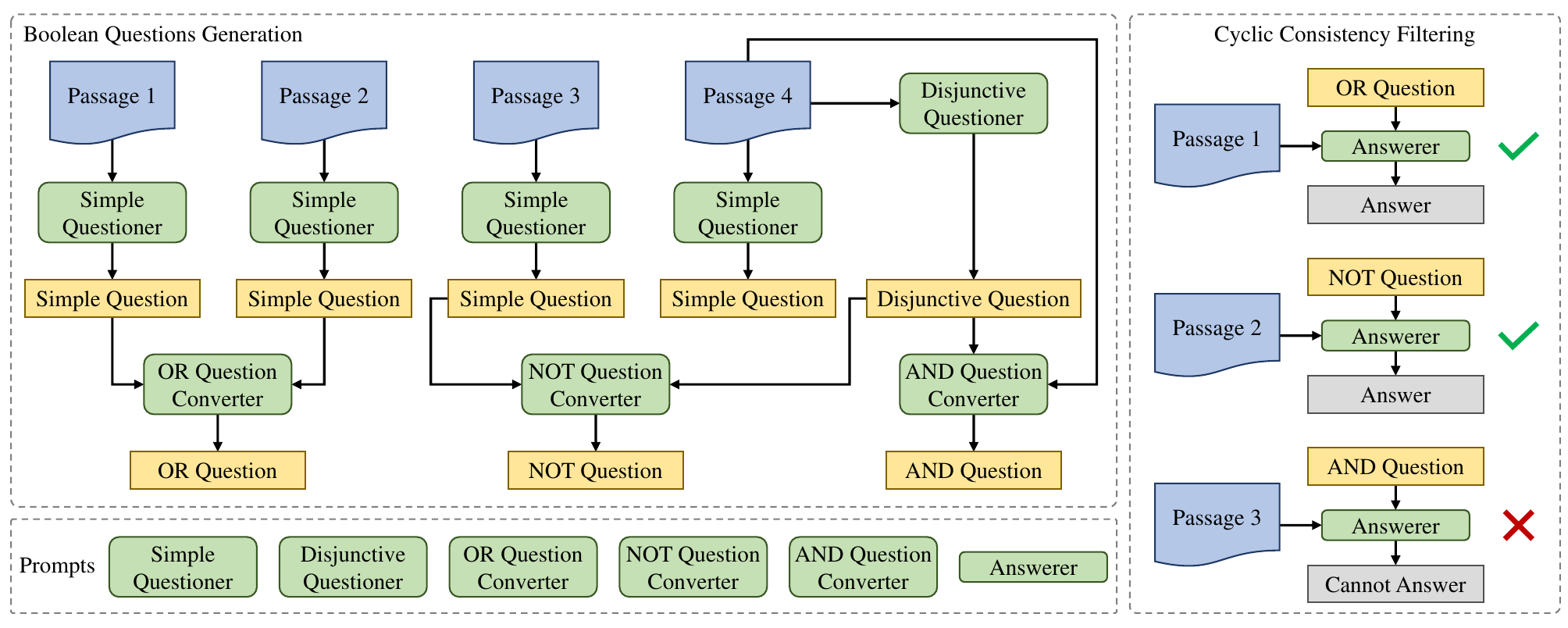}
  \caption{Data collection pipeline of \datasetname.}
  \label{fig:pipeline}
\end{figure*}

\section{Data Collection}

In our work, we developed a question generation framework that progresses from simple to complex structures to gather questions embedded with Boolean logic. Specifically, the process involves sampling text passages from a corpus, generating basic questions from these passages, and then combining these questions using Boolean operators. The combined questions are subsequently rephrased into natural language.  To ensure the relevance between generated questions and corresponding passages, we introduce a cyclic consistency filtering strategy, retaining only those questions that meet this criterion.

Addressing the challenge posed by restricted access to query logs in widely used search engines, we leverage GPT-4\footnote{We use the {gpt-4-0125-preview} version.}, one of the most advanced language generation models currently available. This model is tasked with constructing Boolean questions tailored to the document collections found in prevalent retrieval datasets. Next, we will provide a detailed introduction to our data collection pipeline, which is shown in Figure \ref{fig:pipeline}. All prompts used are detailed in Appendix~\ref{app:prompts}.

\subsection{Passage Clustering}
A complex question containing Boolean logic consists of simple questions with interrelated topics. The passage sampling process must be meticulously designed, as combining questions about disparate objects could result in incoherent and peculiar complex questions. To address this, we propose a hierarchical passage sampling strategy for the subsequent question generation process. Specifically, we conduct clustering on passages of the corpus, assuming that passages from the same cluster discuss similar topics.

\datasetname{} is constructed based on the widely-used passage retrieval datasets MS MARCO~\cite{Bajaj2018MS} and Natural Questions~\cite{kwiatkowski-etal-2019-natural}. For the passage collection in MS MARCO, we first encode the passages using a BERT-style encoder pre-trained for the dense retrieval task\footnote{\url{https://huggingface.co/sentence-transformers/msmarco-distilbert-dot-v5}.}. We then reduce the dimensionality of these vector representations via truncated SVD, with random sampling performed beforehand to lower computation costs. Finally, we apply agglomerative clustering from Sklearn\footnote{\url{https://github.com/scikit-learn/scikit-learn}.} library to the compressed vectors. We also conduct clustering with TF-IDF vectorization but find that clusters formed in this manner focus more on word overlap, which is not suitable for selecting passages within the same topics compared with BERT-style encoder vectorization. The corpus of Natural Questions comprises split English Wikipedia pages, allowing similar topic passages to be directly drawn from successive passages within the same Wikipedia pages.

\subsection{Atomic Question Generation}

For each cluster, we randomly select $n$ passages for atomic question generation. Since a larger number of candidate passages incurs higher computational costs, we randomly set $n$ to 2 or 3. These selected passages are referred to as candidate passages $\{p_i\}_{i=1}^n$.

\paragraph{Simple Question} We prompt GPT-4 to generate simple questions $\{q_i^{\text{simple}}\}_{i=1}^n$ for each candidate passage sampled from the same cluster. Naturally, candidate passages provide answers to their corresponding generated simple questions. Therefore, each simple question is paired with 1 positive passage, from which it is generated.

\paragraph{Disjunctive Question} GPT-4 is also required to generate a disjunctive question $q^{\text{disj}}$ that can be answered with any of the candidate passages individually. The disjunctive question is regarded as a more abstract question than a simple question. Any of the candidate passages provide answers to the disjunctive question and are annotated as positive passages for the generated question.


\subsection{Boolean Question Generation}

Simple questions and disjunctive questions operate at different levels of abstraction. Generally, simple questions focus on more detailed objects than disjunctive questions within the same cluster. We leverage this property to generate complex questions containing Boolean logic.

\paragraph{AND Questions} Disjunctive questions encompass the topics discussed in all of their candidate passages, whereas simple questions within the same cluster address more concrete details of the topic. To simulate the AND operator, we prompt GPT-4 to add extra constraints to the disjunctive question, thus forming an AND question. This can be represented by the Boolean expression:
\begin{equation}
    q^{\text{AND}} \iff q^{\text{disj}}\ [\text{AND}] \ \text{constraints}.
\end{equation}
We ensure that the generated question can be answered solely with a randomly selected candidate passage $\{p^{\text{AND}+}\}$. All remaining candidate passages in the same cluster are labeled as negative passages $\{p_i^{\text{AND}-}\}_{i=1}^{n-1} = \{p_i\}_{i=1}^n \setminus \{p^{\text{AND}+}\}$.

\paragraph{OR Questions} Logical disjunction tends to describe the union of two topics at the same level of abstraction. We randomly selected two simple questions $q_a^{\text{OR}}$ and $q_b^{\text{OR}}$ from $\{q_i^{\text{simple}}\}_{i=1}^n$, join them with OR operator and ask GPT-4 to paraphrase the Boolean expression to a natural language question. Under these circumstances, the OR question is constructed by
\begin{equation}
    q^{\text{OR}} \iff q_a^{\text{OR}}\ [\text{OR}] \ q_b^{\text{OR}}.
\end{equation}
Candidate passages $p_a^{\text{OR}}$ and $p_b^{\text{OR}}$ corresponding to $q_a^{\text{OR}}$ and $q_b^{\text{OR}}$ are annotated as positive passages $\{p_i^{\text{OR}+}\}_{i=1}^{2} = \{p_a^{\text{OR}},p_b^{\text{OR}}\}$, while other candidate passages in the same cluster are labeled as negative passages $\{p_i^{\text{OR}-}\}_{i=1}^{n-2} = \{p_i\}_{i=1}^n \setminus \{p_i^{\text{OR}+}\}_{i=1}^{2}$.

\paragraph{NOT Questions} In contrast to the construction of AND questions, NOT questions are created by adding an exclusion to the disjunctive question. Here, we adopt the information in simple questions as an exclusion to the disjunctive question. Concretely, we concatenate the disjunctive question and a randomly sampled simple question $q^{\text{NOT}}$ with NOT operator to formulate the Boolean expression of NOT question. Then the expression is fed into GPT-4 to be paraphrased into a NOT question in the form of natural language. The creation of NOT questions is defined as
\begin{equation}
    q^{\text{NOT}} \iff q^{\text{disj}}\ [\text{NOT}] \ q_a^{\text{NOT}}.
\end{equation}
Candidate passage $p^{\text{NOT}}$ corresponding to $q^{\text{NOT}}$ is annotated as negative passages $\{p^{\text{NOT}-}\}$ as it contradicts the negation in $q^{\text{NOT}}$, while other candidate passages in the same cluster are labeled as positive passages $\{p_i^{\text{NOT}+}\}_{i=1}^{n-1} = \{p_i\}_{i=1}^n \setminus \{p_i^{\text{NOT}-}\}$.

\subsection{Cyclic Consistency Filtering}
\label{sec:cyclic_consistency_filtering}

Despite that questions generated by strong language generation models are fluent, coherence is not guaranteed. Inspired by the cyclic consistency in image generation~\cite{CycleGAN2017}, we filter the generated questions by checking the cyclic consistency in question answering. Formally, we ask the language generation model whether the passage contains the answer to questions generated from the passage itself. Only questions that can be answered with their associated passages are deemed to be valid questions. After the cyclic consistency filtering, 1151 and 1258 questions are obtained based on MS MARCO and Natural Questions, respectively.


\section{Data Analysis}

In this section, we detail statistics of the proposed \datasetname, analyze the distribution of question types and display examples to provide a more intuitive understanding of our data.

\begin{table}[ht]
	\centering
    \small
	\begin{tabular}{ccc}
		\toprule
		Statistics & \makecell{BQ-MARCO} & \makecell{BQ-NQ} \\

        \midrule
        \#questions (ALL) & 1151 & 1258 \\
        avg \#pos (ALL) & 1.27 & 1.32 \\
        avg \#neg (ALL) & 0.63 & 0.38 \\
        \midrule
        \#questions (AND) & 354 & 403 \\
        avg \#pos (AND) & 1.00 & 1.00 \\
        avg \#neg (AND) & 0.94 & 0.59 \\
        \midrule
        \#questions (OR) & 469 & 485 \\
        avg \#pos (OR) & 1.58 & 1.74 \\
        avg \#neg (OR) & 0.35 & 0.21 \\
        \midrule
        \#questions (NOT) & 328 & 370 \\
        avg \#pos (NOT) & 1.13 & 1.11 \\
        avg \#neg (NOT) & 0.69 & 0.37 \\
        
		\bottomrule
	\end{tabular}
    \caption{Data statistics of \datasetname{} built upon MS MARCO (BQ-MARCO) and Natural Questions~(BQ-NQ).}
	\label{tab:data_statictics}
\end{table}

\begin{table}[ht]
	\centering
    \small
    \renewcommand{\arraystretch}{1.2}
	\begin{tabular}{c|cc}

		\toprule
		& pos (prediction) & neg (prediction) \\
  
        \hline
        pos (ground truth) & 62.50\% & 9.38\% \\
        neg (ground truth) & 3.13\% & 25.00\% \\
		\bottomrule
	\end{tabular}
    \caption{Confusion matrix of 32 random samples in \datasetname{} under human evaluation.}
	\label{tab:confusion_matrix}
\end{table}

\subsection{Data Statistics}
\label{sec:data_statistics}

\begin{table*}[t]
    \vspace{-1em}
    \centering
    \small
    \begin{tabular}{cp{13cm}}
        \toprule
        \makecell[c]{Question Type} & Example(s) \\

        \midrule
        \operator{AND} &
        \textbf{Boolean Question:} How can I start a \keyinfo{career in the accounting field} \operator{and} pursue an \keyinfo{online degree program}?\newline
        \textbf{Paragraph A (Positive):} If you want to work toward a \keyinfo{career in the accounting field}, take the first steps with Penn Foster College. Contact us to learn more about our \keyinfo{online Accounting degree program}. ...\newline
        \textbf{Simple Question for Paragraph A:} Is Penn Foster College's \keyinfo{online Accounting degree program} designed for those aiming to start a career in accounting?\newline
        \textbf{Paragraph B (Negative):} And indeed, if you're speaking primarily to promote your \negkeyinfo{business} ' say, you're offering a seminar on tax preparation ...\newline
        \textbf{Simple Question for Paragraph B:} Do you need to charge for a seminar if promoting your \negkeyinfo{accounting} firm?
        \\
        \midrule

        \operator{OR} &
        \textbf{Boolean Question:} What are the impacts of \keyinfo{global warming} \operator{or} \keyinfo{climate change} on nature and humans?\newline
        \textbf{Paragraph A (Positive):} Global \keyinfo{climate change} will affect people and the environment in many ways. Some of these impacts, like stronger hurricanes and severe heat waves, ...\newline
        \textbf{Simple Question for Paragraph A:} What are the potential impacts of global \keyinfo{climate change} on people and the environment?\newline
        \textbf{Paragraph B (Positive):} \keyinfo{Global warming} is harming the environment in several ways including: 1  Desertification. 2  Increased melting of snow and ice. ...\newline
        \textbf{Simple question for Paragraph B:} What environmental issues are caused by \keyinfo{global warming}?
        \\
        \midrule
        \operator{NOT} &
        \textbf{Boolean Questions:} What causes \keyinfo{upper abdomen pain} \operator{but is unrelated to} \negkeyinfo{liver issues}?\newline
        \textbf{Paragraph A (Positive):} Pain originating in the stomach or esophagus is often felt in the \keyinfo{upper abdomen} and can be due to heartburn, gastroesophageal reflux disease (GERD), or hiatal ...\newline
        \textbf{Simple Question for Paragraph A:} What causes \keyinfo{upper abdomen pain} linked to the stomach or esophagus?\newline
        \textbf{Paragraph B (Negative):} This pain is usually felt in the \keyinfo{upper right part of the abdomen}, often under the rib cage, and is almost always associated with \negkeyinfo{a swelling or enlargement of the liver}, acute inflammation or distention of the liver's surface, or any other ...\newline
        \textbf{Simple Question for Paragraph B:} What symptoms are linked with \negkeyinfo{liver enlargement} or injury?
         \\
        \bottomrule
    \end{tabular}
    \caption{Examples of \datasetname{} built upon MS MARCO corpus. We show in \operator{blue bold} natural language phrases indicating the logical operation, \keyinfo{green bold italics} key information to retrieve the positive passages, and \negkeyinfo{red bold italics} key information to exclude the negative passages.}
    \label{tab:detailed_examples}
\end{table*}

We display the number of questions, average number of positives and negatives for each question of \datasetname{} built upon MS MARCO and Natural Questions. Statistics are calculated on whole datasets and subsets of each type of question individually. Notably, 500 questions are initially generated for each type and we only provide statistics for the final datasets which are filtered by cyclic consistency mentioned in Section~\ref{sec:cyclic_consistency_filtering}.

As AND questions are constructed with the logical conjunction of a disjunctive and a simple question, there is only 1 positive passage for each AND question. OR questions are built by disjunction of several simple questions, and thus more than 1 passages are annotated as positives. NOT questions also have more than 1 positive passage on average since they are collected by removing irrelevant passages within a cluster, where left passages are labeled as positives.

\subsection{Data Quality}

This work utilizes generative language models to construct benchmark datasets without human involvement, and thus it is vital to ensure the data quality for more precise evaluation on retrieval models. Several studies~\cite{li-etal-2023-making,zhao-etal-2023-verify,lightman2024lets} have demonstrated that large language models excel at verification tasks compared to generating novel information. Building on this insight, we deem GPT-4 as a robust filter in our data generation process. However, human involvement is critical for the quality assessment of the generated dataset. Therefore, we randomly sample 32 questions from our dataset and manually check the false-negative and false-positive rates of our proposed Cyclic Consistency Filtering. As the confusion matrix in~\ref{tab:confusion_matrix} indicated, the filtering process effectively eliminates most false positives. Meanwhile, discarding false negatives does not significantly compromise the quality of the filtered data.

\subsection{Question Types}

\begin{figure}[ht]
  \centering
 
  \includegraphics[width=0.85\columnwidth]{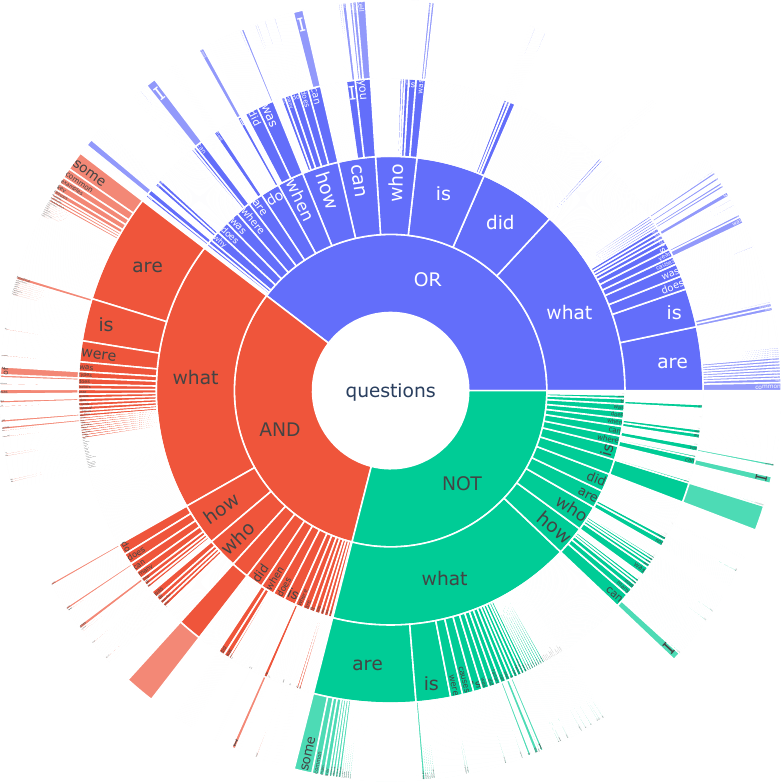}
  \caption{Question types covered in \datasetname. Questions are firstly grouped by the Boolean logic implied in questions and then heuristically categorized following the method in \citet{yang-etal-2018-hotpotqa}. Colored blocks without labels indicate questions whose types can not be determined.}
  \label{fig:question_sunburst}
\end{figure}

Following the approach in~\citet{yang-etal-2018-hotpotqa}, we heuristically analyze the question types of \datasetname, visualized in Figure~\ref{fig:question_sunburst}. \datasetname{} covers a variety of question types including ``what'', ``how'', ``who'', yes/no questions and so on. General questions with leading ``did'' or ``is'' constitute a larger proportion in OR than in other types of questions. One possible explanation is that general questions have a large capacity for questions comprising the disjunction of multiple atomic questions.

\subsection{Data Examples}

We present examples of \datasetname{} in Table~\ref{tab:detailed_examples}. AND and OR questions are easy to create as ``and'' and ``or'' serve as conjunctions in natural language. The main merit of leveraging GPT-4 to generate Boolean questions lies in NOT questions. As the example shown, GPT-4 paraphrases the NOT operator to ``but is unrelated to'' to form a more natural sentence while preserves the Boolean logic. Overall, the quality of generated questions is ensured in our data collection framework.

\section{Experiments}
In this section, we begin by describing the metrics employed in our experiments. We then evaluate the performance of prevalent dense retrievers on the proposed dataset, introduce two baseline methods for approaching the Boolean dense retrieval task, and finally, conduct a detailed analysis of these baseline methods.

\subsection{Metrics for Boolean Dense Retrieval}

Following previous works~\cite{qu-etal-2021-rocketqa,zhang2022adversarial,ren-etal-2021-rocketqav2}, we adopt Mean Reciprocal Rank at top-$k$~(MRR@$k$), which measures the average reciprocal rank of the first retrieved relevant passage, as the main metric of retrieval model performance in our experiments.

In addition, we propose an additional metric named Negative Recall at top-$k$~(NegRecall@$k$) to evaluate the capability of retrieval models for tackling the logic of negation. Formally, NegRecall@$k$ computes the recall of explicit negative passages for the top-$k$ returned retrieval results. Lower NegRecall@$k$ indicates that the retrieval system excludes the related passages in the negation more successfully.

\begin{table*}[!t]
	\centering
    \small
	\begin{tabular}{cp{2.95cm}ccccccccc}
		\toprule
		\multirow{2.5}{1cm}{Corpus} & \multirow{2.5}{1.2cm}{Model} & MRR@10 $\uparrow$ & \multicolumn{4}{c}{\taskname{} MRR@10 $\uparrow$} & \multicolumn{4}{c}{\taskname{} NegRecall@10 $\downarrow$} \\
		\cmidrule(lr){3-3} \cmidrule(lr){4-7} \cmidrule(lr){8-11}
		& & ALL & ALL & AND & OR & NOT & ALL & AND & OR & NOT \\

        \midrule
        \multirow{10}{1cm}{MS MARCO}& distilbert-cos-v5 & 33.78 & 29.11 & 31.46 & 39.44 & 11.80 & \textbf{11.64} & \textbf{3.02} & \textbf{0.00} & 33.08 \\
        & MiniLM-L12-cos-v5 & 32.75 & 31.00 & 32.32 & 43.56 & 11.60 & 11.88 & 3.77 & \textbf{0.00} & \textbf{32.82} \\
        & MiniLM-L6-cos-v5 & 32.25 & 31.09 & 33.80 & 43.98 & 9.73 & 13.08 & 4.15 & \textbf{0.00} & 36.15 \\
        & e5-base & 36.26 & 36.68 & 41.91 & 50.60 & 11.14 & 12.92 & 3.58 & \textbf{0.00} & 36.41 \\
        & distilbert-dot-v5 & 37.25 & 37.61 & 39.92 & 53.04 & 13.04 & 19.50 & 5.28 & 0.61 & 54.62 \\
        & bert-base-dot-v5 & 38.08 & 39.89 & 44.36 & 55.31 & 13.01 & 17.01 & 4.15 & \textbf{0.00} & 48.72 \\
        & distilbert-base-tas-b & 34.43 & 40.00 & 46.60 & 54.48 & 12.18 & 19.98 & 5.28 & \textbf{0.00} & 56.67 \\
        & e5-large-v2 & 35.74 & 41.20 & 49.66 & 54.84 & 12.56 & 16.37 & 3.58 & \textbf{0.00} & 47.44 \\
        & bge-large-en-v1.5 & 35.73 & 42.47 & 46.50 & \textbf{57.90} & 16.06 & 15.25 & 4.34 & \textbf{0.00} & 42.82 \\
        & gte-Qwen2-7B-instruct & \textbf{39.20} & \textbf{43.52} & \textbf{52.01} & 55.54 & \textbf{17.16} & 15.01 & 5.28 & \textbf{0.00} & 40.77 \\
        \midrule
        \multirow{7}{1cm}{Wikipedia}& distilbert-base-v1 & 52.75 & 28.89 & 28.32 & 37.91 & 17.71 & \textbf{20.39} & \textbf{15.84} & \textbf{14.42} & \textbf{32.42} \\
        & dpr-single-nq-base & 56.81 & 35.20 & 36.18 & 43.07 & 23.81 & 35.14 & 34.16 & 23.08 & 46.48 \\
        & stella-en-1.5B-v5 & 44.51 & 42.37 & 42.38 & 51.40 & 30.51 & 36.98 & 33.66 & 22.12 & 54.30 \\
        & bge-large-en-v1.5 & 57.54 & 64.98 & 67.47 & 78.34 & 44.77 & 61.29 & 55.94 & 39.42 & 87.50 \\
        & e5-base-v2 & 64.00 & 67.04 & 72.16 & 80.82 & 43.42 & 61.87 & 55.20 & 45.19 & 85.94 \\
        & e5-large-v2 & 65.93 & 68.61 & 74.20 & \textbf{83.48} & 43.04 & 66.01 & 60.64 & 48.08 & 89.06 \\
        & gte-Qwen2-7B-instruct & \textbf{66.64} & \textbf{71.55} & \textbf{75.65} & 83.03 & \textbf{52.04} & 70.85 & 66.34 & 56.73 & 89.45 \\

		\bottomrule
	\end{tabular}
    \caption{Performance of strong dense retrieval models on the original dataset and \datasetname. MRR@10 without BDR prefix shows the model performance on the original dataset, while BDR MRR@10 and BDR NegRecall@10 denote the model performance on \datasetname. The performance of the whole dataset and subset for specific type of questions are shown individually.}
	\label{tab:main_results}
\end{table*}

\subsection{Performance of Existing Dense Retrieval}

We evaluate several strong dense retrievers on \datasetname, including those with BERT-style architecture and those fine-tuned from large language models that are capable of solving various natural language tasks.

From the results shown in Table~\ref{tab:main_results} we can observe that higher retrieval performance on the original datasets indicates higher performance on our generated \datasetname. Notably, weaker retrievers like distilbert-base-v1 perform better on the original datasets than \datasetname{}, while stronger retrievers like gte-Qwen2-7B-instruct show better performance on \datasetname{} than the original datasets. However, stronger retrievers also suffer from high NegRecall on \datasetname, indicating that these retrievers return more passages possibly relevant to the queries but do not have the ability to distinguish the true positives. 


Among the performance of AND, OR and NOT subset of \datasetname, OR subset enjoys the best performance, achieving 8.9\% higher MRR@10 than AND subset on average. It can also be explained by the average number of positives. OR Questions own more positive passages than other subsets owing to the nature of logical disjunction in the construction of OR questions.

The most remarkable results lie in the NOT subset of \datasetname. MRR@10 for NOT subset of \datasetname{} is significantly worse than MRR@10 for other subsets. On the corpus of MS MARCO, most negative passages recalled are from the NOT questions in consideration of the limited NegRecall@10 on AND and OR subsets. Furthermore, even those retrievers fine-tuned from large language models that are believed to have strong language skills show the same trends with the traditional retrievers. 

In view of these observations mentioned above, we draw the conclusion that logical negation is a challenging problem for current dense retrieval models. This aligns with our intuition that dense retrieval can only model the relevance of texts, but lack the capability of realizing the 
explicit irrelevance. We propose two baseline methods to tackle this problem, detailed in the next section.

\begin{table*}[!t]
    \vspace{-1em}
	\centering
    \small
	\begin{tabular}{ccccccccc}
		\toprule
		\multirow{2.6}{*}{Model} & \multicolumn{4}{c}{\taskname{} MRR@10 $\uparrow$} & \multicolumn{4}{c}{\taskname{} NegRecall@10 $\downarrow$} \\
		\cmidrule(lr){2-5} \cmidrule(lr){6-9}
		& ALL & AND & OR & NOT & ALL & AND & OR & NOT \\

        \midrule
        \multicolumn{1}{l}{distilbert-dot-v5} & \textbf{37.61} & \textbf{39.92} & \textbf{53.04} & 13.04 & 19.50 & 5.28 & 0.61 & 54.62 \\
        \multicolumn{1}{l}{Decomposed Query} & 32.34 & 26.73 & 51.55 & 10.93 & \textbf{2.73} & 3.21 & 0.61 & \textbf{3.85} \\
        \multicolumn{1}{l}{Boolean Contrastive (AND)} & 35.98 & 37.44 & 51.49 & 12.24 & 10.87 & \textbf{2.41} & \textbf{0.00} & 31.44 \\
        \multicolumn{1}{l}{Boolean Contrastive (OR)} & 37.23 & 39.07 & 52.92 & 12.83 & 11.29 & 2.50 & \textbf{0.00} & 32.65 \\
        \multicolumn{1}{l}{Boolean Contrastive (NOT)} & 35.96 & 36.89 & 50.81 & \textbf{13.71} & 13.72 & 3.77 & \textbf{0.00} & 38.72 \\

		\bottomrule
	\end{tabular}
    \caption{Performance comparison of baselines on \datasetname. ``Decomposed Query'' and ``Boolean Contrastive'' denote the Boolean operation on decomposed query and Boolean contrastive continuous training on corresponding train set, respectively. The performance of the whole dataset and each subset are shown individually.}
	\label{tab:baseline_results} 
\end{table*}

\subsection{Baseline Methods}

We propose two baseline methods as a starting point for tackling the Boolean logic implied in language queries:
\begin{itemize}
\item \textbf{Boolean Operation on Decomposed Query} Inverted index is directly incorporated into the dense retrieval procedure in this baseline method. We ask the large language model to decompose complex Boolean questions into a Boolean expression of simple questions. For each simple question, we retrieve top-$2k$ relevant passages as the candidate retrieval results. Then candidate lists of these simple questions are merged based on the Boolean operator in the Boolean expression. Specifically, set intersection, union and difference are performed under AND, OR and NOT operations, respectively. Scores of candidates are recomputed by addition, maxing and subtraction in these situations. Finally, the merged candidate list is ranked according to the recomputed scores.
\item \textbf{Boolean Contrastive Continuous Training} Following the data-driven schema of dense retrieval models, we propose to conduct continued training on pre-trained dense retrievers. We generate 2000 extra training data for AND, OR and NOT questions. These additional data are mixed with original training data to individually fine-tune the pre-trained model using the same objectives in pre-training. For reproducibility, we only conduct continued training on the distilbert-dot-v5 model whose training codes are publicly available.
\end{itemize}

\subsection{Analysis}

We implement the two baselines with distilbert-dot-v5, and show the results in Table~\ref{tab:baseline_results}. It can be observed from the results that both baselines reduce the NegRecall@$k$ on all data samples. Boolean operation on decomposed query reduces the NegRecall@$10$ from 19.50\% to 2.73\%, nearly eliminating the false positives in the Boolean dense retrieval. However, MRR@10 also suffers from significant drops when conducting Boolean operation on decomposed queries. MRR@10 on the AND subset and the whole dataset drops 13.19\% and 5.27\%, respectively, indicating that many true positives are also rejected by this baseline. This is reasonable since there are various ways to decompose a complex question and the logic of decomposed questions and original questions may be not consistent. Na\"ive set subtraction on the retrieval results of decomposed questions exposes the low quality of question decomposition, resulting in low retrieval performance.

Boolean contrastive training on NOT questions impacts the pre-trained model more slightly, reducing MRR@10 from 37.61\% to 35.96\% and NegRecall@10 from 19.50\% to 13.72\% on the whole dataset. Interestingly, Boolean contrastive continuous training improves the MRR@10 and reduces NegRecall@10 simultaneously on the NOT subset of \datasetname, illustrating the improved ability to understand the logic of negation in natural language. Fine-tuning with AND and OR questions does not significantly improve MRR@10 but reduces NegRecall@10. A possible explanation is that AND and OR questions are relatively easier and fine-tuning on extra training set do not lead to further improvement but harm the performance. In contrast, NOT questions are more challenging for current retrievers, thus fine-tuning with them enhances their ability to tackle such questions. We expect that a better-designed learning-based approach would lead to more robust retrieval systems to address the Boolean dense retrieval task.

These results demonstrate that a dilemma exists in retrieving relevant passages for Boolean questions. While NOT questions are the main type of Boolean questions that are hard to tackle, over-emphasizing them could lead to the exclusion of true positive passages in the returned list. 

\section{Conclusions}

In this work, we formulate the task of Boolean dense retrieval and investigate whether current dense retrieval systems understand Boolean logic implied in natural language queries. By collecting a benchmark dataset \datasetname{} and evaluating the performance of prevalent dense retrievers, we find that NOT questions are challenging for current dense retrieval systems to understand correctly. Further, we explore Boolean operation on decomposed query and propose a contrastive continual training method that serves as a strong baseline for the research community.

\section*{Limitations}
Dataset collected in this work is generated by large language models. We have not conduct human annotation or filtering on the generated dataset. Due to limited budget, the number of samples are not large enough to provide a complete training dataset. The diversity and coverage of proposed dataset are limited by the MS MARCO and Natural Questions dataset since we build our dataset based on their corpus. Besides, we only focus on English, evaluations on other languages is limited.

\section*{Acknowledgements}

This work is supported by National Key R\&D Program of China under Contract 2022ZD0119802, National Natural Science Foundation of China under Contract 623B2097 and the Youth Innovation Promotion Association CAS. It was supported by GPU cluster built by MCC Lab of Information Science and Technology Institution, USTC, and the Supercomputing Center of the USTC. This work was also supported by Ant Group Research Fund.

\bibliography{anthology,custom}

\appendix


\clearpage

\section{Prompts}
\label{app:prompts}

\subsection{System Message for Questioner:}
\begin{quote}
    \small
	\begin{tcolorbox}[size=normal,opacityfill=0.05]
	\begin{em}
		You are an experienced questioner and retrieval system tester. You need to generate questions based on the given paragraphs and related instructions, which will be used as queries to test if the retrieval system can understand the Boolean logic contained in natural language. The questions you pose should align as closely as possible with the retrieval system's scenario, meaning the language style of the questions should resemble that of a search engine user. Besides, please vary your expressions more and avoid sticking to just a few ways of saying things. Note, you only need to output one question no longer than 32 words, without any extra content.
	\end{em}
	\end{tcolorbox}
\end{quote}

\subsection{System Message for Answerer:}
\begin{quote}
\small
	\begin{tcolorbox}[size=normal,opacityfill=0.05]
	\begin{em}
		You are an expert answerer who needs to provide answers to the questions based on the given paragraphs. If the question can be answered by the paragraph(s), please provide a brief answer. If the question cannot be answered by the paragraph(s), please respond with "Cannot answer". Note, you only need to output one answer no longer than 64 words or "Cannot answer", without any extra content.
	\end{em}
	\end{tcolorbox}
\end{quote}

\subsection{Prompt for Simple Questioner:}
\begin{quote}
\small
	\begin{tcolorbox}[size=normal,opacityfill=0.05]
	\begin{em}
		Please propose a question that can be answered by the following paragraph.\\
        \\
        $[$PARAGRAPH$]$
	\end{em}
	\end{tcolorbox}
\end{quote}

\subsection{Prompt for Disjuntive Questioner:}
\begin{quote}
\small
	\begin{tcolorbox}[size=normal,opacityfill=0.05]
	\begin{em}
    Please propose a question that can be answered by any of the following paragraphs. Please make sure that each paragraph can provide answers to the question individually.\\
    \\
    $[$PARAGRAPH$]$
	\end{em}
	\end{tcolorbox}
\end{quote}

\subsection{Prompt for AND Question Converter:}
\begin{quote}
\small
	\begin{tcolorbox}[size=normal,opacityfill=0.05]
	\begin{em}
    I need to test whether the retrieval system can understand the logical conjunction (AND) implied in natural language. Please generate a new question by adding constraints to the question "[QUESTION]", so that only paragraphs marked with [positive] provide the answer to the new question, while paragraphs marked with [negative] cannot provide the answer.\\
    \\
    $[$POSITIVE PARAGRAPHS$]$\\
    \\
    $[$NEGATIVE PARAGRAPHS$]$
	\end{em}
	\end{tcolorbox}
\end{quote}

\subsection{Prompt for OR Question Converter:}
\begin{quote}
\small
	\begin{tcolorbox}[size=normal,opacityfill=0.05]
	\begin{em}
    I need to test if the retrieval system can understand the logical disjunction (OR) implied in natural language. Please convert the following expression containing the logical disjunction (OR) into a natural language question.\\
    \\
    $[$LOGICAL EXPRESSION$]$
	\end{em}
	\end{tcolorbox}
\end{quote}

\subsection{Prompt for NOT Question Converter:}
\begin{quote}
\small
	\begin{tcolorbox}[size=normal,opacityfill=0.05]
	\begin{em}
    I need to test if the retrieval system can understand the logic of negation (NOT) implied in natural language. Please convert the following expression containing the logic of negation (NOT) into a natural language question.\\
    \\
    $[$LOGICAL EXPRESSION$]$
	\end{em}
	\end{tcolorbox}
\end{quote}

\subsection{Prompt for Answerer:}
\begin{quote}
\small
	\begin{tcolorbox}[size=normal,opacityfill=0.05]
	\begin{em}
    Please provide a brief answer to the following question according to the given paragraph(s). If the question cannot be answered by the paragraph(s), please respond with "Cannot answer".\\
    \\
    question:\\
    $[$QUESTION$]$\\
    \\
    paragraphs:\\
    $[$PARAGRAPHS$]$
	\end{em}
	\end{tcolorbox}
\end{quote}

\end{document}